\documentclass[twocolumn,aps,prd,nofootinbib,superscriptaddress,showpacs,floatfix,nofootinbib,amsmath,amssymb]{revtex4}

\usepackage{indentfirst}  
\usepackage{bm}    
\usepackage{graphicx}  
\usepackage{amsmath}  
\usepackage{txfonts}
\usepackage{multirow}
\usepackage{upgreek}

\begin{document}

\title{
Exploring positron characteristics utilizing two new positron-electron correlation schemes based on
multiple electronic-structure calculation methods
}

\author{Wen-Shuai Zhang} 
\email{wszhang@mail.ustc.edu.cn} 
\affiliation{Department of Modern Physics, University of Science and Technology of China, Hefei 230026, China} 
\affiliation{State Key Laboratory of Particle Detection and Electronics, USTC, Hefei 230026, China}

\author{Bing-Chuan Gu} 
\email{glacierg@mail.ustc.edu.cn} 
\affiliation{Department of Modern Physics, University of Science and Technology of China, Hefei 230026, China}
\affiliation{State Key Laboratory of Particle Detection and Electronics, USTC, Hefei 230026, China}

\author{Xiao-Xi Han} 
\email{ xiaoxi@mail.ustc.edu.cn } 
\affiliation{Department of Modern Physics, University of Science and Technology of China, Hefei 230026, China} 
\affiliation{State Key Laboratory of Particle Detection and Electronics, USTC, Hefei 230026, China}

\author{Jian-Dang Liu}
\email{liujd@mail.ustc.edu.cn} 
\affiliation{Department of Modern Physics, University of Science and Technology of China, Hefei 230026, China}
\affiliation{State Key Laboratory of Particle Detection and Electronics, USTC, Hefei 230026, China}

\author{Bang-Jiao Ye}
\email{bjye@ustc.edu.cn} 
\affiliation{Department of Modern Physics, University of Science and Technology of China, Hefei 230026, China}
\affiliation{State Key Laboratory of Particle Detection and Electronics, USTC, Hefei 230026, China}

\begin{abstract}
We make a gradient correction to a new local density approximation form of positron-electron correlation. Then the positron lifetimes and affinities are probed by using these two approximation forms based on three electronic-structure calculation methods including the full-potential linearized augmented plane wave (FLAPW) plus local orbitals approach, the atomic superposition (ATSUP) approach and the projector augmented wave (PAW) approach. The differences between calculated lifetimes using the FLAPW and ATSUP methods are clearly interpreted in the view of positron and electron transfers. We further find that a well implemented PAW method can give near-perfect agreement on both the positron lifetimes and affinities with the FLAPW method, and the competitiveness of the ATSUP method against the FLAPW/PAW method is reduced within the best calculations. By comparing with experimental data, the new introduced gradient corrected correlation form is proved competitive for positron lifetime and affinity calculations.
\end{abstract}


\maketitle

\footnotetext{Project supported by
National Natural Science Foundation of China (Grant Nos. 11175171 and 11105139).}

\section{Introduction}  

In recent decades, the Positron Annihilation Spectroscopy (PAS) has become
a valuable method to study the microscopic structure of solids \cite{Tuomi13V85,Yuan2014V31,Li2014V31}
and gives detailed information on the electron density
and/or momentum distribution \cite{Makko14V89} in the regions scanned by positrons.
An accompanying theory is required for a thorough understanding of experimental results.
A full two-component self-consistent scheme \cite{Niemi85V32,Puska95V52}
has been developed for calculating positron states in solids
based on the density functional theory (DFT) \cite{Kohn65V140}.
Especially in bulk material where the positron is delocalized
and does not affect the electron states,
the full two-component scheme can be reduced without losing accuracy
to the conventional scheme \cite{Niemi85V32,Puska95V52}
in which the electronic-structure is determined by common one-component formalism.
However, there are various kinds of approximations
can be adjusted within this calculations.
To improve the analyses of experimental data, one should
find out which approximations are more credible to produce the positron state \cite{Kurip14V89,Kurip14V505,Zhang2015V105} .
In this short paper, we focus on probing the positron lifetimes and affinities
by using two new positron-electron correlation schemes based on
three electronic-structure calculation methods.

Recently, N. D. Drummond \textit{et al.} \cite{Drumm11V107,Drumm10V82}
made two calculations for a positron immersed in a homogeneous electron gas,
by using the Quantum Monte Carlo (QMC) method
and a modified one-component DFT method,
and then two forms of local density approximations (LDA)
on the positron-electron correlation are derived.
Kuriplach and Barbiellini \cite{Kurip14V89,Kurip14V505} proposed a fitted LDA form
and a generalized gradient approximation (GGA) form
based on previous QMC calculation,
and then applied these two forms to multiple calculations
for positron characteristics in solid.
However, the LDA form based on the modified one-component DFT calculation
has not been studied.
In this work, we make a gradient correction to the IDFTLDA form
and validate these two new positron-electron correlation schemes
by applying them to multiple positron lifetimes and affinities calculations.

Besides, we probe in detail the effect of different electronic-structure
calculation methods on positron characteristics in solid.
These methods include
the full-potential linearized augmented plane-wave (FLAPW)
plus local orbitals method \cite{Sjoes00V114},
the projector augmented wave (PAW) method\cite{Bloec94V50},
and the atomic superposition (ATSUP) method \cite{Puska83V13}.
Among these methods,
the FLAPW method is regarded as the most accurate method to calculate electronic-structure,
the ATSUP method performs with the best computational efficiency,
the PAW method has greater computational efficiency and close accuracy as the FLAPW method but
has not been completely tested on positron lifetimes and affinities calculations
except some individual calculations \cite{Wikto14V89,Wikto14V90,Makko06V73,Rauch11V84}.
Moreover, our previous work \cite{Huang14V63} showed that the calculated lifetimes
utilizing the PAW method disagree with that utilizing the FLAPW method.
However, within those PAW calculations, the ionic potential was not well constructed.
In this paper, we investigated the influences of the ionic pseudo-potential/full-potential
and different electron-electron exchange-correlations approaches
within the PAW calculations.
Especially, the difference between calculated lifetimes
by using the self-consistent (FLAPW) and non-self-consistent (ATSUP) methods
is clearly investigated in the view of positron and electron transfers.

This paper is organized as follows:
In Sec. 2, we give a brief and overall description of the models considered here
as well as the computational details and the analysis methods we used.
In Sec. 3, we introduce the experimental data on positron lifetime used in this work.
In Sec. 4, we firstly apply all approximation methods
for electronic-structure and positron-state calculations
to the cases of Si and Al, and give detailed analyses
on the effects of these different approaches,
and then assess the two new correlation schemes by using the positron lifetime/affinity data
in comparison with other schemes base on different electronic-structure calculation methods.

\section{Theory and methodology}

\subsection{Theory} \label{subsec-theory}

In this section, we briefly introduce the calculation scheme for the
positron state and various appproximations investigated in this work.
Firstly, we do the electronic-structure calculation
without considering the perturbation by positron
to obtain the ground-state electronic density $n_{e- }(\vec{r})$
and Coulomb potential $V_{Coul}(\vec{r})$ sensed by positron.
Then, the positron density is determined by solving the Kohn-Sham Eq.:
\begin{equation}
[-\frac{1}{2}\nabla_{\vec{r}}+V_{Coul}(\vec{r})+V_{corr}(\vec{r})]\psi^{+}=
\varepsilon^{+}\psi^{+}, n_{e+ }(\vec{r})=|\psi^{+}(\vec{r})|^{2},
\label{eq:KS}
\end{equation}
where $V_{corr}(\vec{r})$ is the correlation potential between electron
and positron. Finally, the positron lifetime can be obtained by the
inverse of the annihilation rate, which is proportional to the product
of positron density and electron density accompanied by the so-called
enhancement factor arising from the correlation energy between a positron
and electrons \cite{Boron86V34}. The equations are written as follows:
\begin{equation}
\label{eq:tau}
\tau_{e+ }=\frac{1}{\lambda},\ \ \lambda=
\pi r_{0}^{2}c\int d\vec{r}n_{e- }(\vec{r})n_{e+ }(\vec{r})\gamma(n_{e- }),
\end{equation}
where $r_{0}$ is the classical electron radius, $c$ is the speed
of light, and $\gamma(n_{e- })$ is the enhancement factor of the
electron density at the position $\vec{r}$.
The positron affinity can be calculated by adding electron and positron chemical potentials together:
\begin{equation}
\label{eq:affinity}
A^{+}=\mu^{-} + \mu^{+}.
\end{equation}
The positron chemical potential $\mu^{+}$ is determined by the positron ground-state energy.
The electron chemical potential $\mu^{-}$ is derived from the Fermi energy (top energy of the valence band)
in the case of a metal (a semiconductor).
This scheme is still accurate for a perfect lattice,
as in this case the positron density is delocalized
and vanishingly small at every point
thus does not affect the bulk electronic-structure \cite{Boron86V34,Puska95V52}.

\begin{table}[htb]
\centering
\footnotesize
\caption{
\label{tab:enhfct}
Parameterized LDA/GGA correlation schemes.
}
\begin{tabular*}{0.48\textwidth}{@{\extracolsep{\fill}}  l r r r r r r }
\hline
$\gamma$ & $a_2$ 	& $a_3$   & $a_{3/2}$ & $a_{7/3}$ & $a_{8/3}$ & $\alpha$  \\
\hline
\\[-3ex]
IDFTLDA  & 4.1698  &  0.1737 & -1.567 & -3.579 & 0.8364 & 0 	 \\
IDFTGGA  & 4.1698  &  0.1737 & -1.567 & -3.579 & 0.8364 & 0.143 \\
fQMCLDA  & -0.22    &  1/6    & 0 		 & 0 	  & 0 	     & 0  	 \\
fQMCGGA  & -0.22    &  1/6    & 0 		 & 0 	  & 0 	     & 0.05  \\
PHCLDA   & -0.137   &  1/6    & 0 		 & 0 	  & 0 	     & 0   	 \\
PHCGGA   & -0.137   &  1/6    & 0 		 & 0 	  & 0 	     & 0.10  \\
\hline
\end{tabular*}
\end{table}

In practice of this work, each enhancement factor is applied identically to
all electrons as suggested by K. O. Jensen \cite{Jense89V1}.
These enhancement factors can be divided into two categories:
the local density approximation (LDA) and the generalized gradient approximation (GGA),
and parameterized by the following equation,
\begin{eqnarray}
\gamma =
& 1 + (1.23r_{s} + a_{2}r_{s}^{2} + a_3 r_{s}^{3}  + a_{3/2} r_{s}^{3/2} \nonumber \\
& + a_{7/3}r_{s}^{7/3} + a_{8/3}r_{s}^{8/3} )
e^{-\alpha\epsilon} ,
\end{eqnarray}
here, $r_{s}$ is defined by $r_{s}=(3/4\pi n_{e-})^{1/3}$,
$\epsilon$ is defined by
$\epsilon=|\nabla\ln(n_{e- })|^{2} / q_{\textrm{TF}}^{2}$
($q_{\textrm{TF}}^{-1}$ is the local Thomas-Fermi screening length),
$a_{2}$ , $a_3$, $a_{3/2}$, $a_{5/2}$, $a_{7/3}$, $a_{8/3}$ and $\alpha$
are fitted parameters.
We investigated five forms of the enhancement factor and correlation potential
marked by IDFTLDA \cite{Drumm10V82},
fQMCLDA \cite{Kurip14V89,Kurip14V505}, fQMCGGA \cite{Kurip14V89,Kurip14V505},
PHCLDA \cite{Stach93V48} and PHCGGA \cite{Boron10V55},
plus a new GGA form IDFTGGA introduced in this work based on the IDFTLDA scheme.
The fitted parameters of these enhancement factors are listed in Table \ref{tab:enhfct} .
The LDA forms of $V_{corr}$ corresponding to IDFTLDA, fQMCLDA, PHCLDA
are given in Refs. \cite{Drumm10V82}, \cite{Kurip14V89} and \cite{Boron98V57}, respectively.
Within the GGA, the corresponding correlation potential takes the form
$V_{corr}^{ \textrm{GGA} }=V_{corr}^{\textrm{LDA} }e^{-\alpha\epsilon/3}$ \cite{Barbi95V51,Barbi96V53}. 
The electronic density and Coulomb potential
were calculated by using various methods including:
a) the all-electron full potential linearized augmented plane wave plus
local orbitals (FLAPW) method \cite{Sjoes00V114} as implemented in Ref.\cite{Kurip14V89}
being regarded as the most accurate method to calculate electronic-structure,
b) the projector augmented wave (PAW) method \cite{Bloec94V50} with reconstruction of
all-electron and full-potential performing with greater computational efficiency and
close accuracy as the FLAPW method,
c) the non-self-consistent atomic superposition (ATSUP) method \cite{Puska83V13}
performing with the best computational efficiency.

\subsection{Computational details}

During the calculations for electronic-structure, three methods mentioned above 
are implemented in this work.
For FLAPW calculations, the WIEN2k code \cite{Blaha01Voo} was used, the
PBE-GGA approach \cite{Perde96V77} was adopted for electron-electron exchange-correlations,
the total number of k-points in the whole Brillouin zone (BZ) was set to 3375,
and the self-consistency was achieved up to
both levels of 0.0001 Ry for total energy and 0.001 e for charge distance.
For PAW calculations,
the PWSCF code within the Quantum ESPRESSO package \cite{Giann09V21} was used,
the PBEsol-GGA \cite{Perde08V100} and PZ-LDA \cite{Perde81V23} approaches
were also implemented for electron-electron exchange-correlations
besides the PBE-GGA approach,
the PAW pseudo-potential files named \textit{PSLibrary 0.3.1}
and generated by A. D. Corso (SISSA, Italy) were employed \cite{Corso14V95},
the k-points grid was automatically generated with
the parameter being set at least (333) in Monkhorst-Pack scheme,
the kinetic energy cut-off of more than 100 Ry (400 Ry) for the wave-functions
(charge density) and the default convergence threshold of $10^{-6}$
were adopted for self-consistency.
For ATSUP calculations, the electron density and Coulomb potential for each material
were simply approximated by the superposition of the electron density
and Coulomb potential of neutral free atoms \cite{Puska83V13},
while the total number of the node points was set to the same as in PAW calculations.
Besides, the $2 \times 2 \times 2$ supercells were used to calculate the
electron structures of monovacancy in Al and Si.
To obtain the positron-state, the three-dimensional Kohn-Sham equation Eq. (\ref{eq:KS})
was solved by the finite-difference method while the unit cell of each material was divided into about 10
mesh spaces per $bohr$ in each dimension.
All important variable parameters were checked carefully to achieve that the
computational precision of lifetime and affinities are the order of 0.1 ps and 0.01eV, respectively.

\subsection{Model comparison} \label{subsec:ModelComp}

An appropriate criterion must be chosen to make a comparison between different models.
The root mean squared deviation (RMSD) is the most popular one
and defined as the square root of the mean of the squared deviation between
experimental and theoretical results:
$\textrm{RMSD} =   [\sum^{N}_{i=1}
(X_{i}^{\textbf{exp} }-X_{i}^{\textbf{theo} })^{2} / N]^{1/2} $,
here $N$ denotes the number of experimental values.
In addition, since the theoretical values can be treated to be noise-free,
the simple mean-absolute-deviation (MAD) defined by
$\textrm{MAD} =  \sum^{N}_{i=1} [
|X_{i}^{\textbf{modelA} }-X_{i}^{\textbf{modelB} }| / N] $
is much more meaningful to quantify
the overall differences between calculated results by using various models.
It is obvious that the experimental data favor models
producing lower values of the RMSD.

\section{Experimental data}

\begin{table}[!htb]
\centering
\caption{ \label{exp-data}
The experimental values of lifetime $\tau_{\textbf{exp} }$,
the related mean value ${\tau}_{\textbf{exp} }^{*}$
and the corresponding standard deviation $\sigma_{\textbf{exp} }$
for each material involved in this work.
}
%
\footnotesize
\begin{tabular*}{0.48\textwidth}{@{\extracolsep{\fill}}  l r rr }
\hline
\\[-3ex]
 Material & $\tau_{\textbf{exp} }$ & ${\tau}_{\textbf{exp} }^{*}$  & $\sigma_{\textbf{exp} }$  \\
\hline
\\[-3ex]
 Si 	 & 216.7\cite{Campi03V213-215} 218\cite{Campi03V213-215} 218\cite{Campi03V213-215} 222\cite{Campi03V213-215} 216\cite{Campi03V213-215} 	& 218.1 & 2.323   \\
 Ge 	 & 220.5\cite{Campi03V213-215} 230\cite{Campi03V213-215} 230\cite{Campi03V213-215} 228\cite{Campi03V213-215} 228\cite{Campi03V213-215} 	& 227.3 & 3.931   \\
 Mg 	 & 225\cite{SeegeooVoo} 225\cite{Campi03V213-215} 220\cite{Campi03V213-215} 238\cite{Campi03V213-215} 235\cite{Campi03V213-215} 		& 228.6 & 7.569   \\
 Al 	 & 160.7\cite{Campi03V213-215} 166\cite{Campi03V213-215} 163\cite{Campi03V213-215} 165\cite{Campi03V213-215} 165\cite{Campi03V213-215} 	& 163.9 & 2.114   \\
 Ti 	 & 147\cite{SeegeooVoo} 154\cite{Campi03V213-215} 145\cite{Campi03V213-215} 152\cite{Campi03V213-215} 143\cite{Campi03V213-215} 		& 148.2 & 4.658   \\
 Fe 	 & 108\cite{Campi03V213-215} 106\cite{Campi03V213-215} 114\cite{Campi03V213-215} 110\cite{Campi03V213-215} 111\cite{Campi03V213-215} 	& 109.8 & 3.033   \\
 Ni 	 & 109.8\cite{Campi03V213-215} 107\cite{Campi03V213-215} 105\cite{Campi03V213-215} 109\cite{Campi03V213-215} 110\cite{Campi03V213-215} 	& 108.2 & 2.127   \\
 Zn 	 & 148\cite{SeegeooVoo} 153\cite{Campi03V213-215} 145\cite{Campi03V213-215} 154\cite{Campi03V213-215} 152\cite{Campi03V213-215} 		& 150.4 & 3.781   \\
 Cu 	 & 110.7\cite{Campi03V213-215} 122\cite{Campi03V213-215} 112\cite{Campi03V213-215} 110\cite{Campi03V213-215} 120\cite{Campi03V213-215} 	& 114.9 & 2.514   \\
 Nb 	 & 119\cite{Campi03V213-215} 120\cite{Campi03V213-215}  122\cite{Campi03V213-215} 122\cite{Campi03V213-215} 125\cite{Campi03V213-215} 	& 121.6 & 2.302   \\
 Mo 	 & 109.5\cite{Campi03V213-215} 103\cite{Campi03V213-215} 118\cite{Campi03V213-215} 114\cite{Campi03V213-215} 104\cite{Campi03V213-215} 	& 109.7 & 6.418   \\
 Ta 	 & 116\cite{SeegeooVoo} 122\cite{Campi03V213-215} 120\cite{Campi03V213-215} 125\cite{Campi03V213-215} 117\cite{Campi03V213-215} 		& 120.0 & 3.674   \\
 Ag 	 & 120\cite{Campi03V213-215} 130\cite{Campi03V213-215} 131\cite{Campi03V213-215} 133\cite{Welch76V77}  131\cite{SeegeooVoo}				& 129.0 & 5.147   \\
 Au 	 & 117\cite{Campi03V213-215} 113\cite{Campi03V213-215} 113\cite{Campi03V213-215} 117\cite{Campi03V213-215} 123\cite{Campi03V213-215} 	& 116.6 & 4.098   \\
 Cd 	 & 175\cite{SeegeooVoo} 184\cite{Campi03V213-215} 167\cite{Campi03V213-215} 172\cite{Campi03V213-215} 186\cite{Campi03V213-215} 		& 176.8 & 8.043   \\
 In 	 & 194.7\cite{Campi03V213-215} 200\cite{Campi03V213-215} 192\cite{Campi03V213-215} 193\cite{Campi03V213-215} 189\cite{Campi03V213-215} 	& 193.7 & 4.066   \\
 Pb 	 & 194\cite{SeegeooVoo} 200\cite{Campi03V213-215} 204\cite{Campi03V213-215} 200\cite{Campi03V213-215} 209\cite{Campi03V213-215} 		& 201.4 & 5.550   \\
 GaAs  	 & 231.6\cite{Wang00V177} 231\cite{Saari91V44} 230\cite{Polit97V55} 232\cite{Dlube87V42} 220\cite{Danne84V30} 							& 228.9 & 5.043   \\
 InP   	 & 241\cite{Belin98V58} 240\cite{Chen98V66} 247\cite{Puska89V39}  242\cite{Dlube86V7} 244\cite{Dlube85V46} 								& 242.8 & 2.775   \\
 ZnO   	 & 153\cite{Mizun04V45} 159\cite{Braue06V74} 158\cite{Uedon03V93} 161\cite{Brunn01V363-365} 171\cite{Tuomi03V91} 						& 160.4 & 6.618   \\
 CdTe  	 & 284\cite{Plaza94V6} 285\cite{Gely-93V80} 285\cite{Peng00V3} 289\cite{Geffr86Voo} 291\cite{Danne82V15} 								& 286.8 & 3.033   \\
\hline
\end{tabular*}
\end{table}

Up to five recent observed values from different literatures and groups
for 21 materials were gathered to compose a reliable experimental data set.
All the experimental values for each material investigated in this work
are collected basically by using the standard suggested in Ref. \cite{Campi07V19}
and listed in Table \ref{exp-data} with their standard deviation.
Furthermore, the materials having less than five
experimental measurements and/or the older experimental data were avoided being adopted.
It is reasonable to suppose that these materials having insufficient and/or unreliable
experimental data would disrupt the comparison between models.
Especially, the measurements for alkali-metals reported before 1975 are
not suggested to be treated seriously \cite{Kurip14V89}.
The deviations of experimental results between different groups
are usually much larger than the statistical errors,
even when just the recent and reliable measurements are considered.
That is, the systematic error is the dominant factor, so that
the sole statistical error is far from enough and not used in this work.
However, the systematic error is difficult to derive from single experimental result.
So in this paper, the average experimental values of each material were
used to assess the positron-electron correlation models,
and the systematic errors are expected to be cancelled as in Ref. \cite{Campi07V19}.
Because the observed values for defect state are insufficient and/or largely scattered,
it is hard to make a clear discussion on the defect state by using these positron-electron correlation models in this short paper.
Thus, except the detailed analyses in the cases of Si and Al
based on three usually applied approaches for electronic-structure calculations,
we mainly focus on testing the correlation models
by using bulk materials' lifetime data and positron-affinity data.
The experimental data of positron affinity are listed in Table \ref{tab:Aff}.

\section{Results and discussion}

\subsection{Detailed analyses in cases of Si and Al}

\begin{figure}[!htb]
\centering
\includegraphics[width=0.48\textwidth]{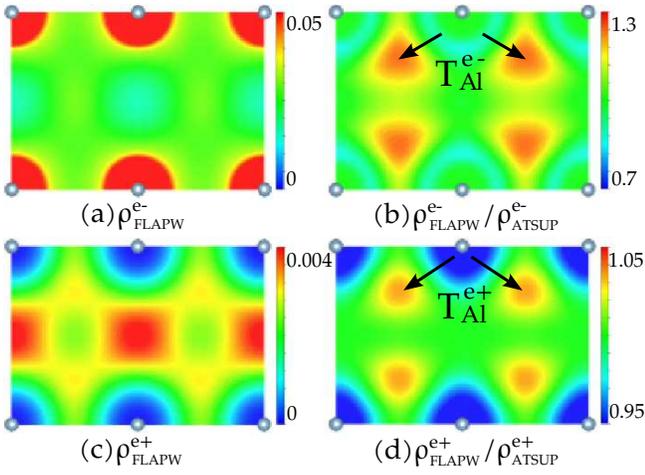}
\caption{
\label{fig:epAl}
Left panels: the self-consistent all-electron (a) and positron densities (c)
(in unit of a.u.) on plane (110) for Al
based on the FLAPW method and the fQMCGGA approximation.
Right panels: the ratios of all-electron (b) or positron densities (d)
calculated by using the FLAPW method to that by using the ATSUP method.
}
\end{figure}

\begin{figure}[!htb]
\centering
\includegraphics[width=0.48\textwidth]{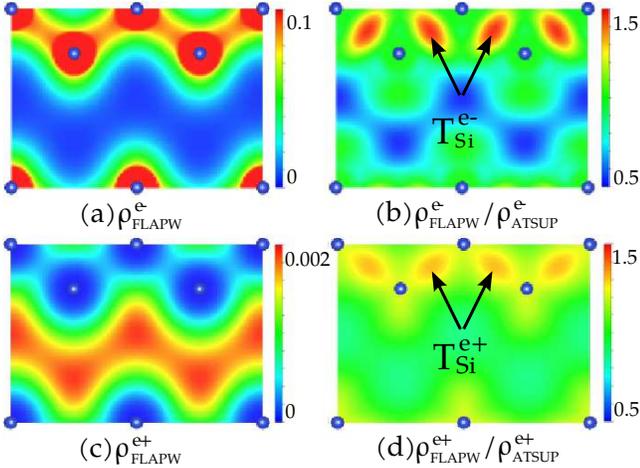}
\caption{
\label{fig:epSi}
As Fig. \ref{fig:epAl}, but for Si.
}
\end{figure}

\begin{figure}[!htb]
\centering
\includegraphics[width=0.48\textwidth]{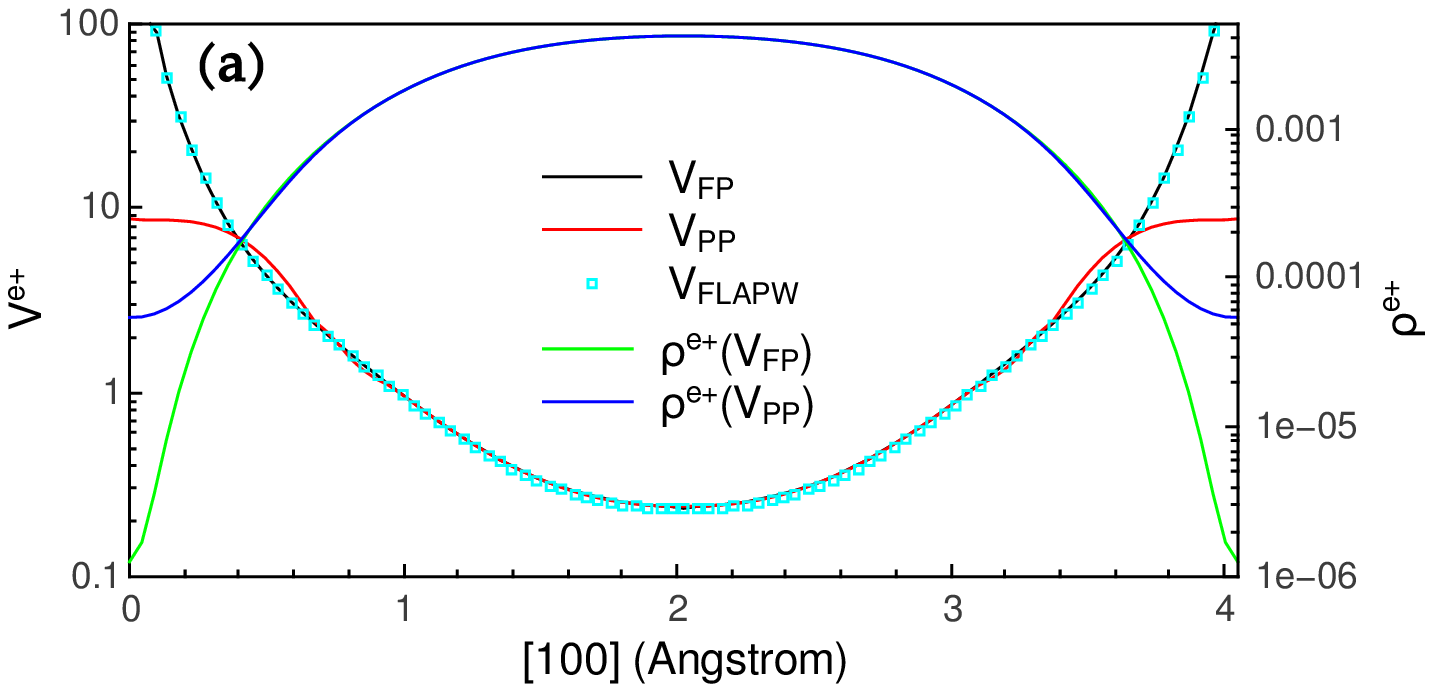}
\includegraphics[width=0.48\textwidth]{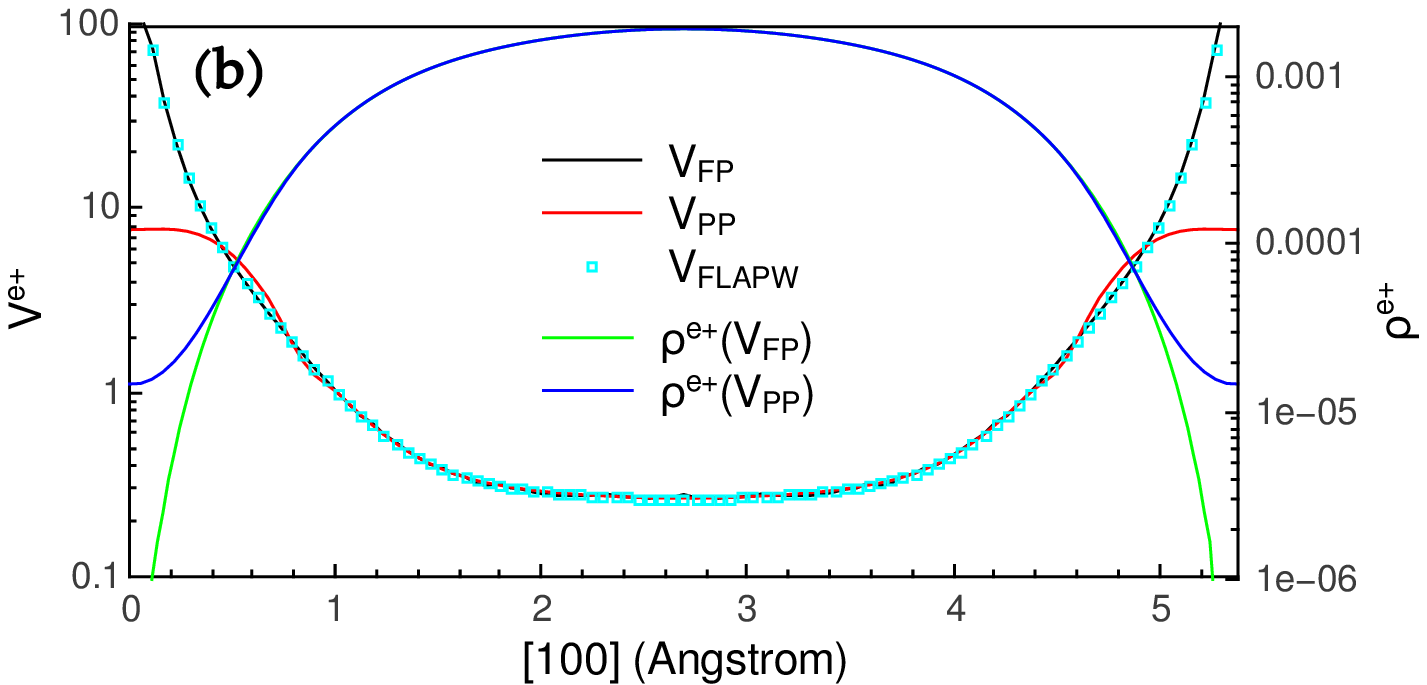}
\caption{
\label{fig:Vcoul-Al-Si}
The total Coulomb potential $\textrm{V}^\textrm{e+}$ (in unit of Ry) sensed by the positron based on 
the ionic pseudo-potentials ($\textrm{V}_\textrm{PP}$)  and
reconstructed ionic full-potential ($\textrm{V}_\textrm{FP}$)
and the corresponding calculated positron densities $\uprho^{\textrm{e+}}$ (in unit of a.u.)
along the [100] direction between two adjacent atoms
for Al (a) and Si (b), respectively.
To make a further comparison,
the full-potentials calculated by using the FLAPW method ($\textrm{V}_\textrm{FLAPW}$) are also plotted.}
\end{figure}

Representatively, the panels (a) and (c) in Fig. \ref{fig:epAl} (Fig. \ref{fig:epSi})
show respectively the self-consistent all-electron and positron densities on plane (110) for Al (Si)
based on the FLAPW method together with
the fQMCGGA form of the enhancement factor and correlation potential.
It is reasonable to obtain that the panel (a) in Fig. \ref{fig:epSi} shows clear
bonding states of Si while the panel (a) in Fig. \ref{fig:epAl} shows
the presence of the nearly free conduction electrons in interstitial regions.
To make a comparison between the FLAPW and ATSUP method for electronic-structure calculations,
we also plot the ratio of their respective all-electron and positron densities
in panel (b) and (d) in Fig. \ref{fig:epAl} (Fig. \ref{fig:epSi}) for Al (Si).
These four ratio panels actually reflect the electron and positron transfers
from densities based on the non-self-consistent free atomic calculations
to that based on the exact self-consistent calculations.
It confirms the fact that the positron density follows the changes of
the electron density which yield not a big difference in annihilation rate
between these two calculations \cite{Puska83V13}.

Now, taking more subtle analyses,
the change of lifetime within the FLAPW calculation
from that within the ATSUP calculation
for Al is attributed to the competition between the following two factors:
a) the lifetime is decreased by the translations of electrons
(illustrated in Fig. \ref{fig:epAl} (b) as $\textbf{T}^{e-}_{\textrm{Al} }$)
from near-nucleus regions with tiny positron densities
to interstitial regions with large positron densities,
b) the lifetime is increased by the translation of positron
(illustrated in Fig. \ref{fig:epAl} (d) as $\textbf{T}^{e+}_{\textrm{Al} }$)
from core regions with large electron densities
to interstitial regions with small electron densities.
However, in the case of Si with bonding states,
the change of lifetime depends conversely on the translations of electrons and positron:
a) the lifetime is increased by the translations of electrons
(illustrated in Fig. \ref{fig:epSi} (b) as $\textbf{T}^{e-}_{\textrm{Si} }$)
from interstitial regions with the largest positron densities
to bonding regions with tiny positron densities,
b) the lifetime is decreased by the translation of positron
(illustrated in Fig. \ref{fig:epSi} (d) as $\textbf{T}^{e+}_{\textrm{Si} }$)
from interstitial regions with tiny electron densities
to bonding regions with large electron densities.
Taking note of the magnitude of scale rulers, these two figures
state clearly that the translations of electrons ($\textbf{T}^{e-}$)
are dominant factors for both Al and Si.
Consequently, the lifetimes within the FLAPW calculations
become smaller (larger) for Al (Si).
These variances are proved by calculated values of lifetimes
listed in Table \ref{tab:LT-Si-Al}.
In addition, the lifetimes of Si calculated by using three GGA forms of the enhancement factor
show greater differences since the large electron-density gradient terms in bonding regions
giving decreases of the enhancement factor
can further weaken the effect of the translation $\textbf{T}^{e+}_{\textrm{Si} }$.

We calculated the bulk lifetimes for Al and Si based on the PAW method.
Within the Table \ref{tab:LT-Si-Al}, the label "PAW" without a suffix indicates that
the electron-structure is calculated by using the PBE-GGA
electron-electron exchange-correlations approach \cite{Perde96V77}
and positron-state is calculated by using reconstructed ionic full-potential (FP),
the suffix "-PZ" indicates that the PBE-GGA approach
is replaced by the PZ-LDA approach \cite{Perde81V23} during electron-structure calculations,
and the suffix "-PP" indicates that ionic full-potential (FP)
is replaced by the ionic pseudo-potential (PP) during positron-state calculations.
The ionic potential together with the Hartree potential from the valence electrons
compose the total Coulomb potential in Eq. (\ref{eq:KS}).
It can be easily found that
the better implemented PAW method by using reconstructed full-potential
can give a startling agreement with the FLAPW method
on the positron-lifetime calculations for Al and Si.
By comparing the results of PAW and PAW-PP approach,
the PAW-PP approach leads to smaller lifetimes
with the differences up to 3.8 ps and 4.3 ps for Al and Si respectively.
These decreases are caused by the fact that the softer potential within the PAW-PP approach
more powerfully attracts positron into the near-nucleus regions with much larger electron densities.
This statement is illustrated by the Fig. \ref{fig:Vcoul-Al-Si} showing
the total Coulomb potential $\textrm{V}^\textrm{e+}$  sensed by the positron based on the 
ionic pseudo-potential ($\textrm{V}_\textrm{PP}$)  and
reconstructed ionic full-potential ($\textrm{V}_\textrm{FP}$)
and the corresponding calculated positron densities $\uprho^{\textrm{e+}}$
along the [100] direction between two adjacent atoms for Al (a) and Si (b), respectively.
To make a further comparison, the full-potentials calculated
by using the FLAPW method ($\textrm{V}_\textrm{FLAPW}$) are also plotted and
found nearly the same as the reconstructed PAW full-potentials.
This figure indicates that a change in the ionic potential approachs (FP or PP)
can lead to a change of more than one order of magnitude in the positron densities near the nuclei.
It should be noted that, in cases of PAW calculations with underestimated core/semicore electron densites
in the near-nucleus regions \cite{Tang2002V65},
the effect of overestimated positron densities based on the pseudo-potentials can be cancelled,
and then excellent quality on the calculated positron lifetimes is able to be achieved.
It is clear that the differences between the results of PAW-PZ and PAW are of the order of 0.1 ps,
and therefore the effect of different electron-electron exchange-correlations schemes is small.
More than this,
we also calculated the lifetimes by using the PBEsol-GGA approach \cite{Perde08V100}
which is revised for solids and their surfaces,
and the similar differences of the order of 0.1 ps are also obtained
compared with the PBE-GGA approach.

\begin{table}[!htb]
\footnotesize
\centering
\caption{
\label{tab:LT-Si-Al}
Calculated results of positron lifetimes (in unit of ps) for
Al, Si, and ideal monovacancy in Al and Si
based on various methods for electronic-structure and positron-state calculations.
}
\begin{tabular*}{0.48\textwidth}{@{\extracolsep{\fill}} p{4mm} l  ccccc  r  }
\hline
 & & IDFT & IDFT & fQMC & fQMC & PHC & PHC  \\
 & & GGA & LDA & GGA & LDA & GGA & LDA  \\
\hline
 \multirow{5}{*}{Al}
& ATSUP   & 160.778 & 152.470 & 173.347 & 169.357 & 163.036 & 156.438 \\
& FLAPW   & 156.615 & 149.852 & 169.972 & 166.530 & 159.397 & 153.878 \\
& PAW     & 156.649 & 149.898 & 170.016 & 166.584 & 159.432 & 153.925 \\
& PAW-PP  & 154.113 & 146.814 & 166.507 & 162.798 & 156.574 & 150.587 \\
& PAW-PZ  & 157.208 & 150.204 & 170.421 & 166.906 & 159.898 & 154.220 \\
\cline{1-1}
\\[-3ex]

 \multirow{5}{*}{Si}
& ATSUP   & 201.770 & 186.634 & 213.260 & 207.345 & 201.363 & 190.484 \\
& FLAPW   & 211.843 & 188.285 & 217.520 & 208.477 & 208.639 & 191.790 \\
& PAW     & 211.779 & 188.245 & 217.466 & 208.431 & 208.586 & 191.752 \\
& PAW-PP  & 208.407 & 184.675 & 213.320 & 204.125 & 205.060 & 187.976 \\
& PAW-PZ  & 211.248 & 188.388 & 217.399 & 208.625 & 208.247 & 191.905 \\
\cline{1-1}
\\[-3ex]

 \multirow{2}{*}{ $\textrm{V}_{\textrm{Al}}$  }
& ATSUP   & 229.441 & 216.639 & 246.294 & 240.941 & 229.686 & 220.274 \\
& PAW     & 212.176 & 201.245 & 229.481 & 224.429 & 214.050 & 205.570 \\
\cline{1-1}
\\[-3ex]

 \multirow{2}{*}{ $\textrm{V}_{\textrm{Si}}$  }
& ATSUP   & 227.458 & 208.972 & 239.524 & 232.309 & 225.922 & 212.690 \\
& PAW     & 236.052 & 208.712 & 241.816 & 231.443 & 231.504 & 212.145 \\
\hline
\end{tabular*}
\end{table}

In addition, as shown in Table \ref{tab:LT-Si-Al},
the positron lifetimes for monovacancy in Al and Si
are also calculated based on the ATSUP and PAW methods for electronic-structure calculations
and six correlation schemes for positron-state calculations.
The ideal monovacancy structure is used in these calculations,
which means that the positron is trapped into a single vacancy
without considering the ionic relaxation from the ideal lattice positions.
Larger differences between the results of ATSUP and PAW are found
in monovacancy-state calculations compared with that in bulk-state calculations.
Besides, the IDFTGGA/IDFTLDA correlation schemes produce similar lifetime values
compared with the PHCGGA/PHCLDA correlation schemes
and produce much smaller lifetime values compared with the fQMCGGA/fQMCLDA correlation schemes
in both monovacancy-state and bulk-state calculations.

\subsection{Positron lifetime calculations}

In this subsection we firstly give visualized comparisons between experimental
values and calculated results based on different methods for
electronic-structure and positron-state calculations.
Within the PAW, the positron lifetimes are all calculated by using the
reconstructed full-potential and certainly all-electron densities from now on.

\begin{figure}[!htb]
\centering
\includegraphics[width=0.48\textwidth]{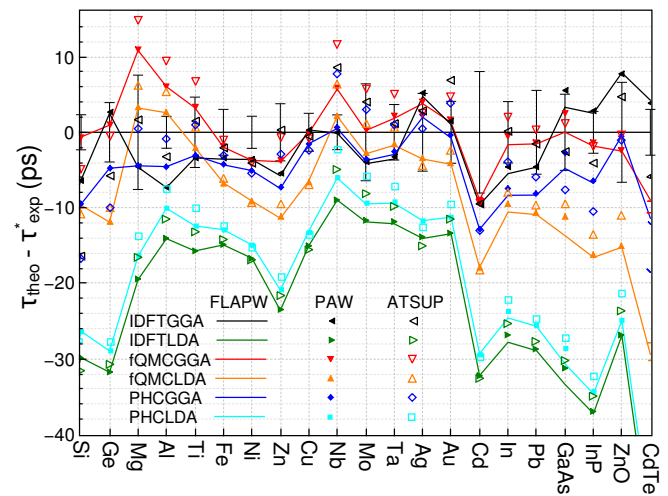}
\caption{
\label{LTs_Allmtds}
The deviations of the theoretical results based on various methods
from the experimental values
alongwith the standard deviation of experimental values for each material.
}
\end{figure}

\begin{table}[!htb]
\centering
\footnotesize
\caption{
\label{tab:RMSD-MAD}
The MADs between the calculated results by using the ATSUP/PAW method and
that by using the FLAPW method. And the RMSDs between the theoretical results and
the experimental data ${\tau}_{\textbf{exp} }^{*}$.
}
\begin{tabular*}{0.48\textwidth}{@{\extracolsep{\fill}} l cccc r  }

\hline
\\[-3ex]
            	& \multicolumn{2}{c}{MAD [ps]} & \multicolumn{3}{c}{RMSD [ps]} 	 \\
\cline{2-3}
\\[-3ex]
            	& ATSUP & PAW      & FLAPW  & PAW    & ATSUP  			 \\
\hline      	                                                           	
\\[-3ex]    	                                                           	
fQMCGGA   		& 2.503 	  & 0.303  		   & 4.503  & 4.591  & 6.309  			 \\
IDFTGGA   		& 5.068 	  & 0.316  		   & 4.809  & 4.821  & 5.611  			 \\
PHCGGA    		& 3.667 	  & 0.287  		   & 6.148  & 6.013  & 7.672  			 \\
fQMCLDA   		& 2.184 	  & 0.290  		   & 11.36  & 11.19  & 10.35  			 \\
IDFTLDA   		& 1.966 	  & 0.253  		   & 25.19  & 24.99  & 23.88  			 \\
PHCLDA    		& 1.936 	  & 0.260  		   & 22.83  & 22.63  & 21.54  			 \\
\hline
\end{tabular*}
\end{table}

The deviations of the theoretical results from the experimental data
alongwith the standard deviations of observed values for all materials
are plotted in Fig. \ref{LTs_Allmtds}.
The scattering regions of calculated results
by different forms of the enhancement factor
are found much larger in the atom systems with bonding states compared with
that in pure metal systems.
Besides, the deviations of the results by using the ATSUP method
from those by using the FLAPW method are mostly larger in GGA approximations
compared with those in LDA approximations.
Numerically, the MADs for different forms of the enhancement factor
between the calculated lifetimes by using the ATSUP method and those by using the FLAPW method
are shown in Table \ref{tab:RMSD-MAD}.
These MADs range from 1.936 ps (PHCLDA) to 5.068 ps (IDFTGGA).
Moreover, the well implemented PAW method is found able to give nearly the same results as the FLAPW method.
Numerically, the MADs between the calculated lifetimes by the PAW method and those
by the FLAPW method for different forms of the enhancement factor
are also shown in Table \ref{tab:RMSD-MAD}.
These MADs range from 0.253 ps (IDFTLDA) to 0.316 ps (IDFTGGA).
This near-perfect agreement between the PAW method and the FLAPW method
proves our calculations are quite credible.

\begin{table*}[!htb]
\footnotesize
\centering
\caption{
\label{tab:Aff}
Theoretical and experimental positron affinities $A^{+}$ (in unit of eV)
based on four positron-electron correlation schemes
and several electron structure calculation methods.
The RMSDs between the theoretical and experimental positron affinities
are also presented.
Here, the PZ-LDA approach is labeled by PZ, and the PBE-LDA approach is labeled by PBE for short.
}
\begin{tabular*}{0.96\textwidth}{@{\extracolsep{\fill}} l rrr rrr rrr rrr  c }
\hline

\multirow{3}{*}{ \ $A^{+}$ }
     & \multicolumn{3}{c}{IDFTGGA} & \multicolumn{3}{c}{IDFTLDA} & \multicolumn{3}{c}{PHCGGA} & \multicolumn{3}{c}{fQMCGGA}                    &   \multirow{3}{*}{ Exp. }   \\
\cline{2-4} \cline{8-10}

     & FLAPW & \multicolumn{2}{c}{PAW} & FLAPW & \multicolumn{2}{c}{PAW} & FLAPW & \multicolumn{2}{c}{PAW} & FLAPW & \multicolumn{2}{c}{PAW} & \\
\cline{3-4} \cline{6-7} \cline{9-10} \cline{12-13}
     & PBE & PBE & PZ & PBE & PBE & PZ & PBE & PBE & PZ & PBE & PBE & PZ &   \\
\\[-3ex]
Si  & -6.481 & -6.478 & -6.683  & -6.884 & -6.881 & -7.070  & -6.728 & -6.726 & -6.926  & -6.182 & -6.179 & -6.373 & -6.2\\
Al  & -4.497 & -4.504 & -4.683  & -4.624 & -4.631 & -4.813  & -4.641 & -4.648 & -4.828  & -3.981 & -3.988 & -4.169 & -4.1\\
Fe  & -3.914 & -3.877 & -4.290  & -4.323 & -4.289 & -4.707  & -4.120 & -4.084 & -4.498  & -3.544 & -3.508 & -3.925 & -3.3\\
Cu  & -4.381 & -4.437 & -4.932  & -4.875 & -4.933 & -5.435  & -4.614 & -4.671 & -5.168  & -4.073 & -4.130 & -4.630 & -4.3\\
Nb  & -3.847 & -3.841 & -4.085  & -4.112 & -4.107 & -4.355  & -4.020 & -4.014 & -4.260  & -3.399 & -3.394 & -3.641 & -3.8\\
Ag  & -5.147 & -5.083 & -5.577  & -5.670 & -5.615 & -6.109  & -5.398 & -5.337 & -5.831  & -4.875 & -4.817 & -5.310 & -5.2\\
W   & -1.956 & -1.982 & -2.304  & -2.225 & -2.254 & -2.580  & -2.121 & -2.149 & -2.472  & -1.491 & -1.520 & -1.844 & -1.9\\
Pb  & -5.954 & -5.936 & -6.305  & -6.328 & -6.305 & -6.683  & -6.186 & -6.166 & -6.538  & -5.622 & -5.601 & -5.977 & -6.1\\
RMSD & 0.285 &  0.283 &  0.546  &  0.570 &  0.566 &  0.899  &  0.431 &  0.427 &  0.740  &  0.314 &  0.314 &	 0.272 &  -  \\
\hline
\end{tabular*}
\end{table*}

Table \ref{tab:RMSD-MAD} also presents the RMSDs between the theoretical results and
the experimental data ${\tau}_{\textbf{exp} }^{*}$
by using six positron-electron correlation schemes.
Two interesting phenomena can be found in this table.
Firstly,
the RMSDs produced by the IDFTLDA scheme
are always worse among the RMSDs based on three electron structure approaches,
but are similar to those produced by the PHCLDA scheme.
Thus, the gradient correction (IDFTGGA) to this LDA form (IDFTLDA) is needed.
It is clear that the corrected IDFTGGA scheme largely improves the calculations,
and performs better than the PHCGGA scheme, but is still worse than the fQMCGGA scheme.
The fQMCGGA scheme together with the FLAPW method produced the best RMSD.
This fact indicates that the quantum Monte Carlo calculation implemented in Ref. \cite{Drumm11V107}
is more credible than the modified one-component DFT calculation \cite{Drumm10V82}
on the positron-electron correlation.
Secondly,
compared to the RMSD produced by using the FLAPW/PAW method,
the RMSD produced by using the simple ATSUP method
is a little smaller based on the LDA correlation schemes,
but is distinctly larger based on the GGA (especially fQMCGGA) correlation schemes.
This phenomenon implies that
the benefit of the exact eletronic-structure calculation approach (PAW/FLAPW)
is swamped by the inaccurate approximation of the enhancement factor.
Meanwhile, the competitiveness of the ATSUP approach against the FLAPW/PAW method
is reduced based on the most accurate positron-electron correlation schemes.

\subsection{Positron affinity calculations}

The positron affinity $A^{+}$ is a important bulk property
which describes the positron energy level in the solid,
and allows us to probe the positron behavior
in an inhomogeneous material.
For example, the difference of the lowest positron energies
between two elemental metals in contact
is given by the positron affinity difference,
and determines how the positron samples near the interface region.
Besides, if the electron work function $\phi^{-} $ is known,
the positron work function $\phi^{+} $ can be derived by the equation:
$\phi^{+} = - \phi^{-} - A^{+} $.
The crystal (e.g., W metal) having a stronger negative positron work function
can emit slow-positron to the vacuum from the surface
and therefore be utilized as a more efficient positron moderator for the slow-positron beam.

The theoretical and experimental positron affinities for eight common materials
by using the new IDFTLDA and IDFTGGA correlation schemes are listed in Table \ref{tab:Aff}.
To make a comparison, the results corresponding to the PHCGGA and fQCMGGA schemes are also listed.
During the electron structure calculation, the ATSUP method was not implemented
because the ATSUP method is inappropriate for positron energetics calculations
and gives much negative positron work functions \cite{Puska83V13}.
Within the PAW calculations, both the PBE-GGA and PZ-LDA approaches are used
for electron-electron exchange-correlations.
The RMSDs between the theoretical and experimental positron affinities
are also presented in Table \ref{tab:Aff}.

As in previous lifetime calculations,
the calculated positron affinities by using the FLAPW method
are also near the same as that by using the PAW method.
Besides, our calculated positron affinities by using the fQMCGGA \& PZ-LDA approaches
are in excellent agreement with that reported in Ref. \cite{Kurip14V89}
with a MAD being 0.06 eV.
Moreover, the differences between the RMSDs produced by using the PBE-GGA and PZ-LDA approaches,
are not negligible,
and the PBE-GGA approach performs mostly better than the PZ-LDA approach
except the case related to fQMCGGA.
In addition, the gradient correction (IDFTGGA) to the IDFTLDA form is needed
to improve the performance for positron affinity calculations.
Meanwhile, the IDFTGGA correlation scheme makes distinct improvement
upon positron affinity calculations compared with the PHCGGA scheme
which is similar to the cases of positron lifetime calculations of bulk materials.
Nevertheless, the best agreement between the calculated and experimental
positron affinities is still given by the fQMCGGA \& PZ-LDA approaches.

\section{Conclusion}

In this work, we probe the positron lifetimes and affinities
utilizing two new positron-electron correlation schemes (IDFTLDA \& IDFTGGA) based on
three common electronic-structure calculation methods (ATSUP \& FLAPW \&PAW).
Firstly, we apply all approximation methods
for electronic-structure and positron-state calculations
to the cases of Si and Al, and give detailed analyses
on the effects of these different approaches.
Especially, the difference between calculated lifetimes
by using the self-consistent (FLAPW) and non-self-consistent (ATSUP) methods
is clearly investigated in the view of positron and electron transfers.
The well implemented PAW method
with reconstruction of all-electron and full-potential,
is found being able to give near-perfect agreement with the FLAPW method,
which proves our calculations are quite credible.
While for ATSUP method, its competitiveness against the FLAPW method is reduced
within calculations utilizing the best positron-electron correlation schemes (fQMCGGA).
Then, we assess the two new positron-electron correlation schemes:
the IDFTLDA form and the IDFTGGA form by using a reliable experimental data on
the positron lifetimes and affinities of bulk materials.
The gradient correction (IDFTGGA) to the IDFTLDA form introduced in this work
is found necessary to promote the positron affinity and/or lifetime calculations.
Moreover, the IDFTGGA performs better than the PHCGGA scheme
in both positron affinity and lifetime calculations.
However, the best agreement between the calculated and experimental
positron lifetimes/affinities is obtained
by using the fQMCGGA positron-electron correlation scheme.
Nevertheless, the new introduced gradient corrected correlation form (IDFTGGA)
is currently competitive for positron lifetime and affinity calculations.

\section*{Acknowledgment}

We would like to thank Han Rong-Dian, Li Jun and Huang Shi-Juan for helpful discussions.
Part of the numerical calculations in this paper were completed on the supercomputing system
in the Supercomputing Center of University of Science and Technology of China.



\begin{thebibliography}{99}


\bibitem{Tuomi13V85}
Tuomisto F and  Makkonen I
\newblock 2013 {\it Rev. Mod. Phys.} {\bf 85} 1583


\bibitem{Yuan2014V31}
Yuan D Q, Zheng Y N, Zuo Y and et~al.
2014 {\it Chin. Phys. Lett.} {\bf 31}(04) 46101

\bibitem{Li2014V31}
Li Y F, Shen T L, Gao X and et~al.
2014 {\it Chin. Phys. Lett.} {\bf 31}(03) 36101


\bibitem{Makko14V89}
Makkonen I, Ervasti M~M, Siro T and  Harju A
\newblock 2014 {\it Phys. Rev.} B {\bf 89} (4) 041105


\bibitem{Niemi85V32}
Nieminen R~M, Boro\'{n}ski E and  Lantto L~J
\newblock 1985 {\it Phys. Rev.} B {\bf 32} 1377


\bibitem{Puska95V52}
Puska M~J, Seitsonen A~P and  Nieminen R~M
\newblock 1995 {\it Phys. Rev.} B {\bf 52} 10947


\bibitem{Kohn65V140}
Kohn W and  Sham L~J
\newblock 1965 {\it Phys. Rev.} {\bf 140} A1133



\bibitem{Kurip14V89}
Kuriplach J and  Barbiellini B
\newblock 2014 {\it Phys. Rev.} B {\bf 89} 155111


\bibitem{Kurip14V505}
Kuriplach J and  Barbiellini B
\newblock 2014 {\it J. Phys.: Conf. Ser.} {\bf 505} 012040
%
%

\bibitem{Zhang2015V105}
Zhang W, Gu B, Liu J and Ye B
\newblock 2015 {\it Comput. Mater. Sci.} {\bf 105} 32


\bibitem{Drumm11V107}
Drummond N~D, L\'{o}pez~R\'{i}os P, Needs R~J and  Pickard C~J
\newblock 2011 {\it Phys. Rev. Lett.} {\bf 107} 207402


\bibitem{Drumm10V82}
Drummond N~D, L\'{o}pez~R\'{i}os P, Pickard C~J and  Needs R~J
\newblock 2010 {\it Phys. Rev.} B {\bf 82} 035107


\bibitem{Sjoes00V114}
Sj\"ostedt E, Nordstr\"om L and  Singh D~J
\newblock 2000 {\it Solid State Commun.} {\bf 114} 15


\bibitem{Bloec94V50}
Bl\"{o}chl P~E
\newblock 1994 {\it Phys. Rev.} B {\bf 50} 17953


\bibitem{Puska83V13}
Puska M~J and  Nieminen R~M
\newblock 1983 {\it J. Phys. F: Met. Phys.} {\bf 13} 333


\bibitem{Wikto14V89}
Wiktor J, Kerbiriou X, Jomard G, Esnouf S, Barthe M~F and  Bertolus M
\newblock 2014 {\it Phys. Rev.} B {\bf 89} 155203


\bibitem{Wikto14V90}
Wiktor J, Barthe M~F, Jomard G, Torrent M, Freyss M and  Bertolus M
\newblock 2014 {\it Phys. Rev.} B {\bf 90} 184101


\bibitem{Makko06V73}
Makkonen I, Hakala M and  Puska M~J
\newblock 2006 {\it Phys. Rev.} B {\bf 73} 035103


\bibitem{Rauch11V84}
Rauch C, Makkonen I and  Tuomisto F
\newblock 2011 {\it Phys. Rev.} B {\bf 84} 125201


\bibitem{Huang14V63}
Huang S~J, Zhang W~S, Liu J~D, Zhang J, Li J and  Ye B~J
\newblock 2014 {\it Acta Phys. Sin} {\bf 63} (21) 217804 (in Chinese)


\bibitem{Boron86V34}
Boro\'{n}ski E and  Nieminen R~M
\newblock 1986 {\it Phys. Rev.} B {\bf 34} 3820


\bibitem{Jense89V1}
Jensen K~O
\newblock 1989 {\it J. Phys.: Condens. Matter} {\bf 1} 10595


\bibitem{Stach93V48}
Stachowiak H and  Lach J
\newblock 1993 {\it Phys. Rev.} B {\bf 48} 9828


\bibitem{Boron10V55}
Boro\'{n}ski E
\newblock 2010 {\it Nukleonika} {\bf 55} 9


\bibitem{Boron98V57}
Boro\'nski E and  Stachowiak H
\newblock 1998 {\it Phys. Rev.} B {\bf 57} (11) 6215


\bibitem{Barbi95V51}
Barbiellini B, Puska M~J, Torsti T and  Nieminen R~M
\newblock 1995 {\it Phys. Rev.} B {\bf 51} 7341


\bibitem{Barbi96V53}
Barbiellini B, Puska M~J, Korhonen T, Harju A, Torsti T and  Nieminen R~M
\newblock 1996 {\it Phys. Rev.} B {\bf 53} 16201


\bibitem{Blaha01Voo}
Blaha P, Schwarz K, Madsen G~K~H, Kvasnicka D and  Luitz J
\newblock WIEN2k, An Augmented Plane Wave Plus Local Orbitals Program for
  Calculating Crystal Properties, Vienna University of Technology, Austria.,
  2001


\bibitem{Perde96V77}
Perdew J~P, Burke K and  Ernzerhof M
\newblock 1996 {\it Phys. Rev. Lett.} {\bf 77} 3865


\bibitem{Giann09V21}
Giannozzi P, Baroni S, Bonini N and  et~al.
\newblock 2009 {\it J. Phys.: Condens. Matter} {\bf 21} (39) 395502


\bibitem{Perde08V100}
Perdew J~P, Ruzsinszky A, Csonka G~I, Vydrov O~A, Scuseria G~E, Constantin L~A, Zhou X and  Burke K
\newblock 2008 {\it Phys. Rev. Lett.} {\bf 100} 136406


\bibitem{Perde81V23}
Perdew J~P and  Zunger A
\newblock 1981 {\it Phys. Rev.} B {\bf 23} 5048


\bibitem{Corso14V95}
Corso A~D
\newblock 2014 {\it Computational Materials Science} {\bf 95} 337


\bibitem{Campi03V213-215}
Campillo~Robles J~M and  Plazaola F
\newblock 2003 {\it Defect Diffus. Forum} {\bf 213-215} 141


\bibitem{SeegeooVoo}
Seeger A, Barnhart F and  W B
\newblock in Positron Annihilation edited by Dorikens-Vanpraet L, Dorikens M
  and Segers D (World Scientific, Singapore, 1989) p. 275s; 
  see also  Sterne P A, Kaiser J H, 1991 {\it Phys. Rev.} B {\bf 43} 13892; 
  and Jensen K O 1989 {\it J. Phys.: Condens. Matter} {\bf 1} 10595


\bibitem{Welch76V77}
Welch D~O and  Lynn K~G
\newblock 1976 {\it Phys. Status Solidi B} {\bf 77} 277


\bibitem{Wang00V177}
Wang Z, Wang S~J, Chen Z~Q, Ma L and  Li S
\newblock 2000 {\it Phys. Stat. Sol. (a)} {\bf 177} 341


\bibitem{Saari91V44}
Saarinen K, Hautoj\"arvi P, Lanki P and  Corbel C
\newblock 1991 {\it Phys. Rev.} B {\bf 44} 10585


\bibitem{Polit97V55}
Polity A, Rudolf F, Nagel C, Eichler S and  Krause-Rehberg R
\newblock 1997 {\it Phys. Rev.} B {\bf 55} 10467


\bibitem{Dlube87V42}
Dlubek G, Krause R, Br\"ummer O and  Tittes J
\newblock 1987 {\it Appl. Phys. A: Solids Surf.} {\bf 42} 125


\bibitem{Danne84V30}
Dannefaer S, Hogg B and  Kerr D
\newblock 1984 {\it Phys. Rev.} B {\bf 30} 3355


\bibitem{Belin98V58}
Beling C~D, Deng A~H, Shan Y~Y, Zhao Y~W, Fung S, Sun N~F, Sun T~N and  Chen X~D
\newblock 1998 {\it Phys. Rev.} B {\bf 58} 13648
\unskip.

\bibitem{Chen98V66}
Chen Z~Q, Hu X~W and  Wang S~J
\newblock 1998 {\it Appl. Phys. A: Solids Surf.} {\bf 66} 435


\bibitem{Puska89V39}
Puska M~J, M\"akinen S, Manninen M and  Nieminen R~M
\newblock 1989 {\it Phys. Rev.} B {\bf 39} 7666


\bibitem{Dlube86V7}
Dlubek G and  Br\"ummer O
\newblock 1986 {\it Ann. Phys. (Leipzig)} {\bf 7} 178


\bibitem{Dlube85V46}
Dlubek G, Br\"ummer O, Plazaola F, Hautoj\"arvi P and  Naukkarinen K
\newblock 1985 {\it Appl. Phys. Lett.} {\bf 46} 1136


\bibitem{Mizun04V45}
Mizuno M, Araki H and  Shirai Y
\newblock 2004 {\it Mater Trans} {\bf 45} 1964


\bibitem{Braue06V74}
Brauer G, Anwand W, Skorupa W, Kuriplach J, Melikhova O, Moisson C, Wenckstern
  H, Schmidt H, Lorenz M and  Grundmann M
\newblock 2006 {\it Phys. Rev.} B {\bf 74} 045208


\bibitem{Uedon03V93}
Uedono A, Koida T, Tsukazaki A, Kawasaki M, Chen Z~Q, Chichibu S and  Koinuma H
\newblock 2003 {\it J. Appl. Phys.} {\bf 93} (5) 2481--2485


\bibitem{Brunn01V363-365}
Brunner S, Puff W, Balogh A~G and  Mascher P
\newblock 2001 {\it Mater. Sci. Forum} {\bf 363-365} 141


\bibitem{Tuomi03V91}
Tuomisto F, Ranki V, Saarinen K and  Look D~C
\newblock 2003 {\it Phys. Rev. Lett.} {\bf 91} 205502


\bibitem{Plaza94V6}
Plazaola F, Seitsonen A~P and  Puska M~J
\newblock 1994 {\it J. Phys.: Condens. Matter} {\bf 6} 8809


\bibitem{Gely-93V80}
G\'ely-Sykes C, Corbel C and  Triboulet R
\newblock 1993 {\it Solid State Commun.} {\bf 80} 79


\bibitem{Peng00V3}
Peng Z~L, Simpson P~J and  Maschera P
\newblock 2000 {\it Electrochem. Solid-State Lett.} {\bf 3} (3) 150



\bibitem{Geffr86Voo}
Geffroy B, Corbel C, Stucky M, Triboulet R, Hautoj\"arvi P, Plazaola F~L,
  Saarinen K, Rajainm\"aki H, Aaltonen J, Moser P, Sengupta A and  Pautrat J~L
\newblock 1986 {\it Defects in Semiconductors}, ed. \ H. J. von Bardeleben,
  Materials Science Forum (Trans Tech Publications, Aedermannsdorff, 1986)
  Vols10-12, p1241


\bibitem{Danne82V15}
Dannefaer S
\newblock 1982 {\it J. Phys.} C {\bf 15} 599


\bibitem{Campi07V19}
Campillo~Robles J~M, Ogando E and  Plazaola F
\newblock 2007 {\it J. Phys.: Condes. Matter} {\bf 19} 176222
%
%

\bibitem{Tang2002V65}
Tang Z, Hasegawa M, Nagai Y,  Saito M and  Kawazoe Y
\newblock 2002 {\it Phys. Rev. B} {\bf 65} 045108






\end{thebibliography}
\end{document}